\documentclass[preprint,aps]{revtex4}

\usepackage{graphicx}

\begin{document}

\title{The radiative potential method for calculations of QED radiative corrections 
to energy levels and electromagnetic amplitudes in many-electron atoms}
\author{V.V. Flambaum$^\dagger$ and J.S.M. Ginges$^\dagger$$^\ddagger$}
\affiliation{$^\dagger$School of Physics, University of New South Wales, Sydney 2052, Australia\\
$^\ddagger$Department of Physics, University of Alberta, Edmonton AB T6G 2J1, Canada}
\date{\today}

%************************************************************
\begin{abstract}

We derive an approximate expression for a ``radiative potential'' which can be used to calculate 
QED strong Coulomb field radiative corrections to energies and electric dipole (E1) transition amplitudes 
in many-electron atoms with an accuracy of a few percent. 
The expectation value of the radiative potential gives radiative corrections to the energies. 
Radiative corrections to E1 amplitudes can be expressed in terms of the radiative potential and its 
energy derivative (the low-energy theorem): the relative magnitude of the radiative potential contribution is 
$\sim \alpha^3 Z^2 \ln({1/\alpha^2 Z^2})$, while the sum of other QED contributions is $\sim \alpha^3 (Z_i+1)^2$, 
where $Z_i$ is the ion charge; 
that is, for neutral atoms ($Z_i=0$) the radiative potential contribution exceeds other contributions $\sim Z^2$ times. 
The advantage of the radiative potential method is that it is very simple and can be easily incorporated into 
many-body theory approaches: relativistic Hartree-Fock, configuration interaction, many-body perturbation theory, etc.
As an application we have calculated the radiative corrections to the energy levels and E1 amplitudes  
as well as their contributions (-0.33\% and 0.42\%, respectively) to the parity
non-conserving (PNC) $6s$-$7s$ amplitude in neutral cesium (Z=55). 
Combining these results with the QED correction to the weak matrix elements 
(-0.41\%) we obtain the total QED correction to the PNC $6s$-$7s$ amplitude, (-0.32 $\pm$ 0.03)\%. 
The cesium weak charge $Q_W=-72.66(29)_{\rm exp}(36)_{\rm theor}$ agrees with
the Standard Model value $Q_W^{\rm SM}=-73.19(13)$, the difference is 0.53(48).

\end{abstract}

\maketitle
%************************************************************

\section{Introduction}

The precision of calculations and measurements of phenomena of heavy neutral atoms has 
reached the level where strong-field QED radiative corrections are observable. 
The most striking example is parity nonconservation (PNC) in the neutral cesium atom ($Z=55$) where the nuclear Coulomb field 
radiative corrections ``saved'' the standard model of particle physics 
(this dramatic story may be found, e.g., in the review \cite{FG}; see also the original papers for the measurement 
\cite{wood97,wieman99} and calculations of strong-field radiative corrections 
\cite{johnson01,milstein,dzuba2002,kf_jpb_2002,kf_prl_2002,k_jpb_2002,MST_PRL_2002,kf_jpb_2003,MST_PRA_2003,SPVC_PRA_2003,Shabaev,commentPNC}).

While there is an abundance of highly-accurate calculations of radiative corrections to phenomena of 
single-electron or few-electron atoms, only a handful of calculations have been performed for atoms with many electrons. 
A proper account of the many-body effects in calculations of radiative corrections to phenomena of many-electron 
atoms, including the PNC amplitude, is lacking. 

The first estimates of radiative corrections to energies of an external electron in heavy neutral atoms 
were performed more than 20 years ago \cite{dzuba83}. 
A semi-empirical formula for the $s$-wave radiative correction to energy levels (the Lamb shift) was derived.
The relative magnitude of this correction is $\sim  Z^2\alpha ^3 \ln{(1/Z^2\alpha ^2)}$. 
It rapidly increases with the nuclear charge $Z$ and was important in making a very accurate prediction 
of the francium ($Z=87$) spectrum (measurements performed after the theoretical prediction \cite{dzuba83} agree 
with the calculated energy levels to better than 0.1\%). 
Calculations of the Lamb shift in alkali and coinage metal atoms were performed in Ref. \cite{Labzowsky} using 
local Dirac-Slater potentials; importantly, it was demonstrated that the ratio of energy shifts arising from the Uehling 
potential and self-energy in neutral atoms is the same as that 
in hydrogen-like ions, verifying the authors' earlier estimates \cite{Labzowsky_est}. 
Calculations of the Lamb shift in neutral alkali atoms were also performed in Ref. \cite{Sapirstein_energies} using local 
atomic potentials.
In our recent works \cite{dzuba2002} we used a parametric potential fitted to reproduce radiative corrections in 
hydrogen-like ions to perform approximate numerical calculations of radiative corrections to the energy levels, 
electric dipole (E1) transition amplitudes, and a rough estimate of radiative corrections to the PNC amplitude in cesium. 
Recently, calculations of radiative corrections in local effective atomic potentials have been performed for
E1 amplitudes in neutral alkalis in Ref. \cite{Sapirstein1} and for the PNC amplitude in Cs in Ref. \cite{Shabaev}. 

It is well-known that many-body effects (exchange interaction, core relaxation and polarization, correlations) 
may be very important. 
For any perturbation located at small distances, e.g. the field (volume) isotopic shift, the matrix elements for 
an electron with orbital angular momentum $l>0$ are dominated by the many-body effects mentioned above. 
In this case one cannot guarantee the magnitude or even the sign of the shift when one does a 
model potential calculation. Moreover, even for the $s$-wave, many-body corrections can change
the results by a factor of 2. QED radiative corrections also come from small distances, 
and one may expect a similar situation. Therefore, recent calculations of QED corrections 
in model atomic potentials cannot guarantee results of high accuracy. Indeed, in a 
recent work \cite{uehling_relax} the importance of a proper account of core relaxation in 
calculations of the radiative shift due to the Uehling potential was demonstrated for neutral cesium.

In the present work we suggest a simple radiative potential approach which allows one to calculate 
radiative corrections to many-electron atoms including many-body effects. 
In particular, our method is valid for calculations of the Lamb shift and radiative corrections to E1 amplitudes.
We claim an accuracy of a few percent for calculations of radiative corrections to $s$-wave energy levels, 
$s$-$p$ intervals, and $s$-$p$ E1 amplitudes for neutral cesium using the radiative potential approach. 
We believe that our approach is complementary to the direct Feynman diagram calculations in a model potential.
 
This paper is organized as follows.
In Section \ref{radiative_potential} we derive an approximate ab initio formula for the radiative potential 
for $Z\alpha \ll 1$. Then we refine this potential to include higher orders in $Z\alpha$ using published results 
for hydrogen-like ions \cite{mohr}.
In Section \ref{E1_amplitudes} we describe the procedure for calculations of QED radiative corrections 
to electric dipole amplitudes. In Section \ref{low-energy_theorem} we derive the low-energy theorem:
the vertex and normalization corrections are expressed in terms of $\frac{\partial \Sigma}{\partial E}$
where $\Sigma$ is the electron self-energy operator. Further discussion is presented in Section \ref{E1_discussion}.
The dominating contribution is due to the radiative corrections to electron wave functions produced by
the radiative potential. The relative value of this contribution is 
$\sim \alpha^3 Z^2 \ln({1/\alpha^2 Z^2})$ both in ions and neutral atoms.
It is shown in Section \ref{renorm_and_estimates} that the sum of all other terms 
(the vertex correction, the normalization correction, and part of the self-energy operator 
which can be presented as $\Sigma^A=(H-E)A+A(H-E)$ and does not contribute to the energy shifts) 
is $\sim \alpha^3 (Z_i+1)^2$ where $Z_i$ is the ion charge, $H$ is the atomic electron Hamiltonian, 
and $A$ is an operator defined in Section \ref{renorm_and_estimates}. 
Therefore, in neutral atoms ($Z_i=0$) this sum is $Z^2$ times smaller than the radiative potential 
contribution and may be neglected. 
As an application we calculate in Section \ref{Cs} the radiative corrections to the energy levels and 
E1 amplitudes in neutral cesium including many-body effects: 
core relaxation, core polarization by the photon electric field, and correlation corrections. 
In Section \ref{PNC} we calculate the radiative corrections to energy levels and E1 amplitudes 
contributing to the parity non-conserving $6s$-$7s$ amplitude in Cs. 
Finally, the conclusion for the radiative potential approach for the calculations of the QED radiative corrections 
to energies, E1 amplitudes, and the PNC amplitude is given in Section \ref{conclusion}.

\section{The radiative potential}
\label{radiative_potential}

\subsection{Derivation of the radiative potential; radiative shifts in H-like ions}

\begin{figure}[h]
\centerline{\includegraphics[width=4in]{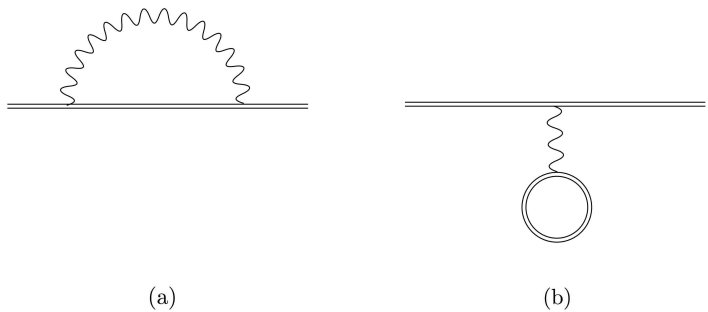}}
\caption{\label{fig:rad_shifts}
Diagrams for radiative energy shifts, $\langle \hat{L} \rangle$, 
corresponding to (a) the self-energy and (b) vacuum polarization. 
Double line denotes a bound electron; wavy line denotes a photon.}
\end{figure}

We define a radiative potential $\hat{L}$ such that its 
average value coincides with the radiative corrections to energies,  
\begin{equation} \label{L}
\delta E_n=  \langle n|\hat{L}|n \rangle \ ;
\end{equation}
the radiative potential is non-local and energy-dependent, 
$\hat{L}=\hat{L}({\bf r},{\bf r}^\prime,E)$, where $E$ is the electron energy.
It contains the non-local electron self-energy operator in the strong Coulomb field, 
$\hat{\Sigma}({\bf r},{\bf r}^\prime,E)$, and the local vacuum polarization operator
comprised of the lowest-order in $Z\alpha$ Uehling potential $\Phi _{U}(r)$ 
and the higher-order Wichmann-Kroll potential. Diagrams for the radiative energy shifts 
are presented in Fig. \ref{fig:rad_shifts}.

The actual problem is the calculation of the self-energy $\Sigma({\bf r},{\bf r}',E)$ contribution 
to the radiative potential. 
This calculation can be divided into two parts: 
one in which the electron interaction with virtual photons of high-frequency is considered, 
and one in which virtual photons of low-frequency is considered.
In the high-frequency case the external field (the nuclear Coulomb field) need only 
be included to first order (vertex diagram).
In the case of a free electron the vertex diagram gives the electric 
$f(q^{2})$ and magnetic $g(q^{2})$ formfactors
presented, for example, in the book \cite{berestetskii}.
The calculations of the contributions of $f(q^{2})$ and $g(q^{2})$ to the radiative potential $\hat{L}$ 
are similar to the calculation of the Uehling contribution presented in \cite{berestetskii}.
Therefore, we present the calculations very briefly. 
We also present the well-known results for the Uehling potential
for comparison with the $f(q^{2})$ and $g(q^{2})$ contributions calculated in this work.

In the momentum representation the high-frequency contribution
to the  radiative potential is equal to 
\begin{equation}\label{Vq}
 \Phi_{\rm rad}({\bf q})=Q_{\rm rad}({\bf q}) \Phi({\bf q}) \ ,
\end{equation}
where $\Phi$ is the atomic potential which at small distances
is equal to the unscreened nuclear electrostatic potential, and 
\begin{equation}\label{Q}
Q_{\rm rad}({\bf q})=-\frac{1}{{\bf q}^2} P(-{\bf q}^2) + \frac{g(-{\bf q}^2)}{2m}\mbox{\boldmath$\gamma$}
\cdot \mbox{\boldmath$q$}
 + f(-{\bf q}^2) -1 \ .
\end{equation} 
Here the first term contains the polarization operator  $P(-{\bf q}^2)$ and
leads to the Uehling potential, $\mbox{\boldmath$\gamma$}$ are the Dirac matrices, 
and $m=1/r_c$ where $r_c$ is the Compton wavelength. We use units $\hbar = c = 1$ throughout, 
except where presented explicitly.
In the coordinate representation \cite{berestetskii} 
\begin{equation}\label{Vr}
\Phi_{\rm rad}(r)=\frac{1}{4 \pi^2 r} {\rm Im} \int^{\infty}_{-\infty} \Phi_{\rm rad}(-y^2)
 \exp(iry) y dy \ ,
\end{equation} 
where $y=\sqrt{{\bf q}^2}$. 
A method to calculate this integral is suggested in \cite{berestetskii}.
 After substitution of $P(-{\bf q}^2)$,  $f(-{\bf q}^2)$ and $g(-{\bf q}^2)$
from  \cite{berestetskii} we obtain
\begin{equation}\label{Phi}
\Phi_{\rm rad}(r)=\Phi_U(r)+\Phi_g(r)+\Phi_f^{\lambda}(r) \ .
\end{equation}
The first term is the well-known Uehling potential
\begin{equation}\label{U1}
\Phi _{U}(r)= \frac{2 \alpha }{3\pi}\Phi (r)
\int _{1}^{\infty}dt~ \frac{\sqrt{t^2-1}}{t^2}
\Big(1 +\frac{1}{2t^2}\Big) 
{\rm e}^{-2trm} \ .   
\end{equation}
For the magnetic formfactor contribution we obtain 
\begin{equation}\label{g}
\Phi _{g}(r)= \frac{ \alpha }{4\pi m} i
\mbox{\boldmath$\gamma$}
\cdot \mbox{\boldmath$\nabla$}\Big[
\Phi (r)\Big(\int _{1}^{\infty}dt~ \frac{1}{t^2 \sqrt{t^2-1}}
{\rm e}^{-2trm} \,\,\, -1 \Big) \Big] \ .
\end{equation}
A straightforward calculation for the electric formfactor gives
\begin{equation}\label{lambda}
\Phi _f^{\lambda}(r)=-\frac{\alpha }{\pi}\Phi (r)
\int _{1}^{\infty}dt~ \frac{1}{\sqrt{t^2-1}}
\Big[\Big(1- \frac{1}{2t^2}\Big) \Big(\ln{(t^2-1)}+\ln{(4 m^2/\lambda^2)}\Big)
-\frac{3}{2}+\frac{1}{t^2}\Big] 
{\rm e}^{-2trm} \ .   
\end{equation}

The expression for the electric formfactor (\ref{lambda}) contains a low-frequency cut-off parameter $\lambda$ 
in the argument of the logarithm. In the standard calculation of the energy shift \cite{berestetskii} 
this parameter is assumed to be in the interval $(Z\alpha)^2 m \ll \lambda \ll m$. After addition of the
 low-frequency contribution the parameter  $\lambda$ cancels out.
To minimize the low-frequency contribution we select $\lambda$ of the order of the electron binding energy 
in atoms, $\lambda \sim (Z\alpha)^2 m$, which is the smallest possible value that can be taken as the border 
for the high-frequency region (one can use free-electron Green's functions to calculate the electron formfactors, 
used above, for frequencies $\omega \gg (Z\alpha)^2 m$).
This  gives $\ln{(4 m^2/\lambda^2)}=4 \ln{(1/Z\alpha)}+const$, where the constant $const$ does not depend on 
$Z$ for $Z\alpha \ll 1$, $ const \sim 1$.
Therefore, the high-frequency contribution allows us to determine the electric formfactor 
contribution with logarithmic accuracy.
The uncertainty is due to the omitted low-frequency contribution for this term. 

The constant in the electric formfactor can be found from a comparison with calculations of the 
Lamb shift for the high-energy levels (principal quantum number $n \gg 1$) in hydrogen-like ions.
Indeed, the energy of an external electron in a many-electron atom is extremely
small, $E \sim 10^{-5} m c^2$.
Therefore, we need the self-energy operator $\Sigma(E\approx 0)$.
The short-range character of $\Sigma({\bf r},{\bf r}',0)$ means that we can
use unscreened Coulomb Green's functions to calculate it.

For very light atoms ($Z \sim 1$) the comparison can be made with the non-relativistic 
calculations presented in the book \cite{berestetskii}, yielding $const \approx -0.63$.
However, this result is not applicable for $Z \ge 10$. 
In this case we have to use the results of all-orders in $Z\alpha$ calculations for hydrogen-like ions 
presented in Ref. \cite{mohr}. To reproduce the results of \cite{mohr} (and leave some room for a low-frequency 
contribution discussed below) we select this logarithm in the form $4\ln{(1/Z\alpha+0.5)}$, where a small constant
$0.5$ is added into the argument of the logarithm. 
This selection gives an accuracy $\sim 10 \%$  for all important applications: 
the $s$-wave self-energy, $s$-$p$ intervals (which are needed to calculate the parity violation effects), 
and fine structure intervals for any $Z$. However, the radiative shifts for $p$-waves (and higher waves) 
are small and sensitive to the low frequency contribution. To make our calculation complete and improve
the accuracy to 1\% we should consider this contribution too.

A consistent calculation of the low-frequency contribution to the non-local self-energy operator 
$\Sigma({\bf r},{\bf r}',E)$ using Coulomb or parametric potential Green's functions is a complicated task. 
However, at the present level of experimental accuracy this low-frequency problem is not of 
immediate importance. It is much easier to fit this small low-frequency contribution using a 
parametric potential $\Phi_l (r)$.
A typical frequency in the low-frequency contribution is $\sim E_{1s}$; 
therefore, the range of this potential is about the size of the $1s$ orbital, $a_B/Z$. 
To reproduce the $p$-level radiative energy shifts we use the following
expression for the low-frequency contribution
\begin{equation}\label{l}
\Phi _{l}(r)= -\frac{B(Z)}{e}Z^4 \alpha^5 m c^2 {\rm e}^{-Zr/a_B} \ ,
\end{equation}
where $e$ is the proton charge and $B(Z)=0.074 + 0.35 Z\alpha$ is a 
coefficient fitted to reproduce the radiative shifts for the high
Coulomb $p$-levels calculated in \cite{mohr}.

Finally, we should introduce one more correction which becomes important
for very heavy atoms, $Z>80$. The potential (\ref{lambda})
 is not applicable for very small distances, $r \ll Z \alpha r_c$.
Indeed, we used an expression for the electric formfactor of a free electron.
However, at very small distances the electron potential energy
$Z \alpha/r \gg m$ and we should use an expression for an off-mass-shell
formfactor $f(p,p')$ instead of $f(-q^2)$. The formfactor 
 $f(p,p')$ leads to a non-local expression $\Sigma({\bf r},{\bf r}',E)$
for $r \ll Z \alpha r_c$ instead of the local potential (\ref{lambda}).
Integration of an electron wave function $\psi$ with a non-local
operator ( $\int \Sigma({\bf r},{\bf r}',E) \psi({\bf r}') d^3 r'$) makes the effective
potential less singular. We take into account this fact by introducing
a small distance cut-off coefficient $mr/(mr + 0.07Z^2\alpha ^2)$.
Our final expression for the electric formfactor contribution
has the following form  
 \begin{equation}\label{f}
\Phi _{f}(r)=-A(Z,r)\frac{\alpha }{\pi}\Phi (r)
\int _{1}^{\infty}dt~ \frac{1}{\sqrt{t^2-1}}
\Big[\Big(1- \frac{1}{2t^2}\Big) \Big(\ln{(t^2-1)}+4\ln{(1/Z\alpha+0.5)}\Big)
-\frac{3}{2}+\frac{1}{t^2}\Big] 
{\rm e}^{-2trm} \ .   
\end{equation}
Here the coefficient  $A(Z,r)=(1.071-1.976x^2-2.128x^3+0.169x^4)
mr/(mr+0.07Z^2\alpha ^2)$, where $x=(Z-80)\alpha$; 
$A(Z,r)$ was found by fitting the radiative shifts for the high
 Coulomb $s$-levels calculated in \cite{mohr}.

Thus, we obtain the following expression for the complete radiative potential
\begin{equation}\label{Phif1}
\Phi_{\rm rad}(r)=\Phi_U(r)+\Phi_g(r)+ \Phi_f(r) + \Phi_l(r)+
\frac{2}{3} \Phi^{\rm simple}_{WC}(r) \ ,
\end{equation}
where $\Phi_U(r)$ is the Uehling potential (\ref{U1}), $\Phi_g(r)$ is the magnetic formfactor contribution 
(\ref{g}), $\Phi_f(r)$ is the high-frequency electric formfactor contribution (\ref{f}), and 
$\Phi_l(r)$ is the low-frequency contribution (\ref{l}).

To make the picture complete we added to the radiative potential
a simplified form of the Wichmann-Kroll potential (higher-orders vacuum polarization) 
which has accurate short-range and long-range asymptotics \cite{dzuba2002}: 
\begin{equation}\label{WC}
\Phi^{\rm simple} _{WC}(r)= -\frac{2 \alpha }{3\pi}\Phi (r)
 \frac{0.092 Z^2\alpha^2}{1+(1.62 r/r_c)^4} \ .   
\end{equation}
Accurate calculations of the Wichmann-Kroll contribution in hydrogen-like ions for $Z=30,40,...100$ 
have been performed in Ref. \cite{Sapirstein}. To reproduce the Wichmann-Kroll
$s$-wave shifts from \cite{Sapirstein}
with a few percent accuracy one should take $\Phi _{WC}(r)=\frac{2}{3}\Phi^{\rm simple} _{WC}(r)$. 
The Wichmann-Kroll potential (\ref{WC}) gives a very 
small contribution which may be noticeable ($\sim 1\%$) only for $Z>80$. 
(This confirms our conclusion \cite{dzuba2002} that the contribution of higher orders into the 
vacuum polarization potential is so small that it is unobservable in neutral atoms: 
for the cesium atom the Wichmann-Kroll potential gives a contribution to $s$-waves about
50 times smaller than the Uehling potential contribution; moreover, the Uehling potential 
itself gives only about 10\% of the total radiative correction for $s$-waves and a very 
small contribution for higher waves.)

Note that our fitting coefficient in the electric formfactor contribution
$A(Z,r) \approx 1$ and the coefficient in the low-frequency contribution $B$ is small, therefore the semi-empirical 
radiative potential (\ref{Phif1}) is always close to the result of the direct Feynman diagram calculation
$\Phi_U(r)+\Phi_g(r)+ \Phi_f^{\lambda}(r)$. 
It may look surprising that the radiative potential obtained in the approximation
$Z\alpha \ll 1$ gives energy shifts which are close to the all-orders results. 
However, the higher-order corrections to the energy shifts are mainly due to the
relativistic electron wave functions which we take into account exactly
when calculating the matrix elements. Indeed, the Dirac
wave function diverges at small distances $r<a_B/Z$:
\begin{equation}\label{psi}
\frac{\psi_{p1/2}^\dagger \psi_{p1/2}}{Z^2 \alpha ^2}
 \sim \psi_s^\dagger \psi_s\sim
 r^{-Z^2 \alpha ^2}
=\exp{(-Z^2 \alpha ^2\ln r)} \ .  
\end{equation}
The radiative energy shifts originate from very small distances $r \sim r_c=\hbar/ mc$.
Thus, we take into account the relativistic enhancement factor 
  $(\psi^2(r_c)/\psi^2(a_B/Z)) \sim (a_B/Zr_c)^{Z^2 \alpha^2}=(1/Z \alpha)^{Z^2 \alpha^2}=
\exp{(Z^2 \alpha ^2\ln{(1/Z \alpha)})}$ in the $Z\alpha$ dependence of the matrix elements
when we use the Dirac wave functions.

\begin{table}[h]
\caption{Difference between the results of the Mohr-Kim self-energy calculations in Ref. \cite{mohr} 
and the radiative potential results ($({\rm Mohr} -g -f -l)/({\rm Mohr}-g)$) in \% for hydrogen-like ions.}
\label{tab:a}
\begin{ruledtabular}
\begin{tabular}{c|ccccccccccc}
Z    & 10    & 20    & 30    & 40    & 50    & 60    & 70    & 80    & 90    & 100   & 110   \\
\hline
$5s_{1/2}$ & 0.0 & 0.4 & 0.5 & 0.3 & 0.0 & -0.2 & -0.2 & 0.0 & 0.1 & 0.1 & 0.0 \\
\hline
$5p_{1/2}$ & -0.8 & -3.6 & -2.8 & -1.8 & -1.1 & -0.7 & -0.3 & 0.1 & 0.8 & 1.8 & 3.3 \\
\hline
$5p_{3/2}$ & -2.5 & -8.3 & -8.9 & -7.3 & -5.2 & -3.1 & -1.1 & 0.4 & 1.4 & 1.7 & 0.8 \\
\end{tabular}
\end{ruledtabular}
\end{table}

In Table \ref{tab:a} we compare the self-energy for $5s$, $5p_{1/2}$, and $5p_{3/2}$ levels 
in hydrogen-like ions calculated using the potential
 $\Phi_g(r)+ \Phi_f(r) + \Phi_l(r)$ 
with those of Ref. \cite{mohr}.  
It is seen that the radiative potential $\Phi _{\rm rad}$ reproduces the self-energy within a few percent 
for all $Z$. 
The comparison is made for the highest available principal quantum number $n=5$ in order to satisfy the condition 
$E \ll m c^2$ which is needed to calculate the radiative corrections in neutral atoms.
However, in practice the results are good for any $n>1$. 
Moreover, the potential $\Phi_{\rm rad}(r)$ even gives the $1s$ energy with reasonable accuracy, $\sim 10\%$.
Note that to calculate parity violation we mainly need to reproduce high $s$-level shifts in 
Cs ($Z=55$), Tl ($Z=81$), and Fr ($Z=87$); % where the accuracy $\sim 0.1\%$. 
the $p$-level shifts are very small and not important.

The above calculations (and those of Ref. \cite{mohr}) were performed in the Coulomb field of a 
point-like nucleus $\Phi (r)=Ze/r$. 
The small correction due to finite nuclear size can be taken into account using integration 
over a realistic charge density for the nucleus, 
$\Phi _{\rm rad}({\bf r})=\int \Phi _{\rm rad}^{\rm point\ charge}(|{\bf r}-{\bf r}'|) \rho({\bf r'}) d^3 r'$.
The finite nuclear size contribution is suppressed by a small parameter $r_n/r_c \sim 10^{-2}$. 
The results of our calculations for the neutral cesium atom presented later in this work include this correction.
 
\subsection{The radiative potential in atomic calculations}

The radiative potential we have derived from radiative shifts in hydrogen-like ions can be used in 
calculations of radiative shifts in ions and neutral atoms for all $Z$ and for any number of electrons.   

Indeed, all electron wave functions with energy $E \ll mc^2$ are proportional to the zero-energy Coulomb
wave functions in the area $r \sim  r_c$, since the energy $E$ may be neglected in the Dirac equation 
and the potential is unscreened in this region.
Therefore, the ratio of the matrix elements of the radiative potential will be proportional
to the ratio of the electron densities near the origin (at a given $j,l$).
This is the reason why one may use parametric potentials fitted to reproduce Lamb-shifts in 
hydrogen-like ions (for principal quantum numbers $n \gg 1$).
Any potential of the range $\sim r_c$ will give  the same results. 
The radiative potentials Eqs. (\ref{U1},\ref{f}) belong to this class.
%In the limit  $r_c \rightarrow 0$ the results coincides with that of \cite{berestetskii}.
This also explains the conclusion of Ref. \cite{Labzowsky} that the ratio of the self-energy 
contribution to the Uehling contribution is the same in hydrogen-like ions and neutral atoms 
calculated in Dirac-Slater potentials.

The magnetic formfactor potential Eq. (\ref{g}) is a long-range one. 
However, it decays rapidly and its matrix elements
are still determined by small distances $r \sim a_{B}/Z$ where all the wave functions with 
principal quantum numbers $n \gg 1$ and given $j,l$ are proportional 
(since the energy $|E_n| \ll |E_{1s}|$ and may be neglected in the area $r \sim a_{B}/Z$). 
Thus, the radiative shifts at a given $j,l$ are still proportional to the electron
density in the vicinity of the nucleus ($1/n^3$ in the Coulomb case). 
Numerical data presented in Ref. \cite{berestetskii} show that this statement
is accurate to a few percent for $n>1$ (exact for $n \gg 1$).

Our semi-empirical radiative potential was derived in the field of the nucleus. 
In atoms there is an electron density contribution to the radiative potential. 
This can be found by integration of the point-charge radiative potential over the electron density 
(as in the finite nuclear size calculation).
It is easy to show that this contribution is very small. 
The electron density contribution is suppressed $Z$ times relative to the nuclear
charge contribution. Indeed, to find the energy shift
 we need to integrate the radiative potential with a squared external electron wave function $\psi^2(r)$. 
For the nuclear contribution  $\psi^2 \sim \psi^2(r_c)$; 
for the electron density contribution $\psi^2 \sim \psi^2(a_B)$, since the radius of the electron
charge density is of the order of the Bohr radius $a_B$.
For an external $s$-wave electron in a neutral atom the ratio $\psi^2(r_c)/\psi^2(a_B) \sim Z$.
Thus, the nuclear contribution is $Z$ times larger than the electron density
contribution. A more elaborate estimate using the WKB semiclassical electron wave function and the 
Thomas-Fermi electron density confirms this simple estimate. 

Note that to estimate the electron density contribution to the self-energy
operator we can use a semiclassical expression for $\Sigma$ derived
in \cite{Zelevinsky}:
\begin{equation}\label{semi}
\Sigma(r,r',E)=\frac{\alpha }{3\pi m^2}\ln{\frac{mc^2}{|e\Phi (r)+E|}}\,\,
\nabla^2 (-e\Phi (r)) \,\,\delta ({\bf r}-{\bf r'}) \ .
\end{equation}
Here $\nabla^2 (-e\Phi (r))= 4 \pi e^2 n_e(r)$, $n_e(r)$ is the electron
number density. This semiclassical expression is valid for $r>a_B/Z$.
Again, an estimate based on Eq. (\ref{semi}) shows that the electron density 
contribution is $Z$ times smaller than the nuclear charge contribution.

Another conclusion from Eq. (\ref{semi}) is that the self-energy operator
is not sensitive to the energy of a valence electron in the area $r \sim a_B/Z$ 
(the small-distance boundary of the applicability of Eq. (\ref{semi})) where
$e\Phi (a_B/Z) \sim (Z^2 \alpha^2) m c^2 \gg E \sim \alpha^2 m c^2$.
The logarithm in this area is equal to $\ln{(1/Z^2 \alpha^2)}$,
it is the same value that appears in the pure Coulomb case (c.f. Eq. (\ref{f})). 
An estimate of the energy dependence in this area may be characterized by the ratio 
\begin{equation}\
\label{semider} 
\frac{\frac{\partial \Sigma}{\partial E}}{(\Sigma/E)} \sim \frac{1}{Z^2}
\end{equation}
which is very small in neutral atoms. In ions this ratio is
 $\sim (Z_i+1)^2/Z^2$, where $Z_i$ is the ion charge
(for a valence electron in an ion $E \sim (Z_i+1)^2\alpha^2 m c^2$).
We shall recall these conclusions during the discussion of the low-energy
theorem for an electric dipole amplitude.

\subsection{Asymptotics of the radiative potential}

It may be useful to present long-range and short-range asymptotics of the radiative potentials. 
For $mr \gg 1$
\begin{equation}\label{Ularge}
\Phi _{U}(r)= \frac{\alpha }{4\sqrt{\pi}(mr)^{3/2}}{\rm e}^{-2mr}\Phi (r) \ ,
\end{equation}
\begin{equation}\label{flarge}
\Phi _{f}(r) \sim \frac{\alpha }{4\sqrt{\pi}(mr)^{1/2}}\Big(\ln{mr}+4\ln{Z\alpha}
\Big){\rm e}^{-2mr}\Phi (r) \ .
\end{equation}  
Note that the asymptotics of the high-frequency contribution to $\Phi _{f}(r)$ are presented 
as an illustration only. A correct (long-range) expression for large $r$ is determined by the 
contribution of low frequencies. However, numerically this long-range contribution is not significant.
%much smaller than that of the magnetic g-factor.
Indeed, the radiative corrections to $s$-wave energies are proportional to 
$1/n^3$ ($\psi^2$ near the origin) to an accuracy $\sim 1\%$ (see \cite{berestetskii,mohr}). 
This can be considered as an estimate of the contribution of the long-range tail and the energy
dependence of $\Sigma({\bf r},{\bf r}',E)$.
%This also follows from the values of the radiative shifts in p-wave presented
%in \cite{berestetskii,mohr}. 

The formfactor $g$ at large distances gives a contribution which describes the interaction of the 
electron anomalous magnetic moment with the atomic electrostatic potential $\Phi$ \cite{berestetskii}:
\begin{equation}\label{glarge}
\Phi _{g}=-i\frac{\alpha \hbar}{4\pi mc}\mbox{\boldmath$\gamma$}
\cdot \mbox{\boldmath$\nabla$}\Phi \ .
\end{equation}
This long-range potential decreases faster than $1/r^2$ since the nuclear electrostatic potential 
is screened by atomic electrons. It gives an especially important contribution for orbitals with $l>0$. 
The long-range character of this interaction guarantees that it is not very sensitive to higher order 
in $Z\alpha$ corrections which are produced by the strong Coulomb field at $r \sim r_c$.

The short-range asymptotics of the radiative potentials, $mr \ll 1$, are the following:  
\begin{equation}\label{Ushort}
\Phi _{U}(r)= \frac{2\alpha }{3\pi}\ln{(1/mr)}\Phi (r) \ ,
\end{equation}
\begin{equation}\label{fshort}
\Phi _{f}(r)= -\frac{\alpha }{\pi}\ln{(1/mr)}
\Big(\ln{(1/mr)}+\ln{(m^2/\lambda^2)}\Big)\Phi (r) \ ,
\end{equation}
\begin{equation}\label{gshort}
\Phi _{g}(r)= i\frac{\alpha}{2\pi}mr\ln{(1/mr)}
\mbox{\boldmath$\gamma$}\cdot \mbox{\boldmath$n$}\Phi (r) \ .
\end{equation}
We see that the area $mr<1$ is not important for the magnetic formfactor contribution.
As we pointed out in the previous section the expression (\ref{fshort}) for the electric
formfactor contribution is not applicable for $mr \ll Z\alpha$. Indeed, the short-range
asymptotics (\ref{fshort}) can be obtained very easily 
using high-energy asymptotics of the vertex operator $\Gamma^{\mu}(p,p',q)$
where $p$ and $p'$ are initial and final electron 4-momenta, and $q$ is the
photon 4-momentum. These asymptotics can be found, e.g., in \cite{berestetskii}.
In the case $|q^2| \gg p^2={p'}^2=m^2$ 
\begin{equation}\label{gamma1}
\Gamma^{\mu}(p,p',q)=\gamma^{\mu}\exp{[-\frac{\alpha}{4 \pi}\ln{\frac{|q^2|}{m^2}}
(\ln{\frac{|q^2|}{m^2}}+2\ln{\frac{m^2}{\lambda^2})}]} \ .
\end{equation}
To obtain the result in the coordinate representation we should substituite $q \sim 1/r$. 
This gives the correction (\ref{fshort}) to the Coulomb potential.
However, in the area  $|q^2| \gg p^2,{p'}^2 \gg m^2$
\begin{equation}\label{gamma2}
\Gamma^{\mu}(p,p',q)=\gamma^{\mu}\exp{[-\frac{\alpha}{2 \pi}\ln{\frac{|q^2|}{p^2}}
\ln{\frac{|q^2|}{{p'}^2}}]} \ .
\end{equation}
Here we have an explicit dependence on both electron momenta, therefore in the coordinate
representation we obtain the non-local correction to the potential.
This corresponds to the area where the Coulomb potential $Z \alpha/r \gg m$.

For electron angular momentum $j>1/2$, 
the short-range contributions of $\Phi _{U}(r)$ and $ \Phi _{f}(r)$ are suppressed.
The dominating contribution is given by the long-range $\Phi _{g}(r)$. 
The result is also sensitive to the low-frequency contribution described by $\Phi _{l}(r)$.
The radiative shift for the $p_{1/2}$ orbital is a special case. 
This orbital has a lower Dirac component penetrating to the region $r \sim r_c$, see Eq. (\ref{psi}). 
The relative contribution of the short-distance area $r \sim r_c$ increases as $\sim Z^2 \alpha^2$. 
This leads to a cancellation of $ \Phi _{f}(r)$ and $ \Phi _{g}(r)$ contributions at $Z \sim 40$. 
For very large $Z$ the short-range contribution dominates and the radiative corrections for $p_{1/2}$ 
become comparable to that for $s_{1/2}$. 
At $Z=110$ the $p_{1/2}$ shift is only 3 times smaller than the $s_{1/2}$ shift.
  
\section{Electromagnetic E1 amplitudes}
\label{E1_amplitudes}

\begin{figure}[h]
\centerline{\includegraphics[width=6in]{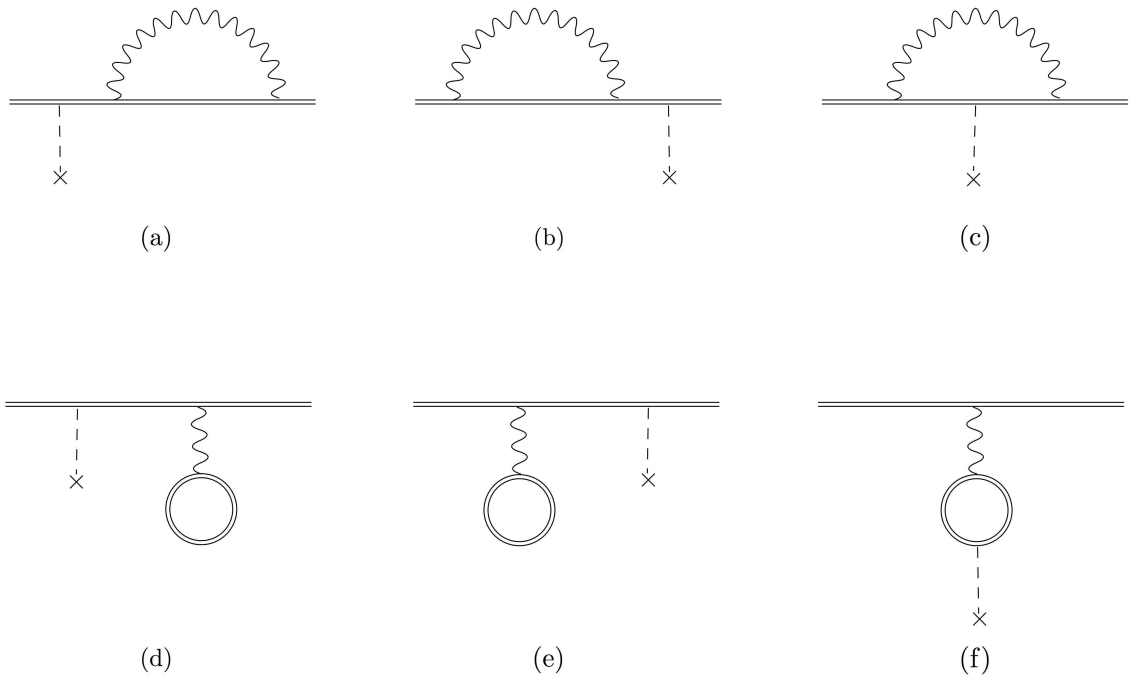}}
\caption{\label{fig:rad_E1}
Diagrams for radiative corrections to the E1 amplitude: 
(a),(b) self-energy; (c) vertex; (d),(e) vacuum polarization; and (f) external photon polarization operator. 
Notations are the same as in Fig. \ref{fig:rad_shifts}; 
dashed line with the cross denotes the laser driven external photon field.}
\end{figure}

Diagrams for radiative corrections to the electric dipole (E1) transition amplitude are presented 
in Fig. \ref{fig:rad_E1}.
The magnitude of different QED contributions to E1 amplitudes depends 
on the virtual photon gauge. Corrections to E1 amplitudes in hydrogen-like ions have been calculated 
with logarithmic accuracy ($\sim \alpha^3 Z^2 \ln{(1/\alpha^2 Z^2)}$) in Ref. \cite{Ivanov}. 
It is pointed out there that in the Yennie gauge it is enough to take into account
only those corrections to the non-relativistic electron wave functions produced   
by the non-relativistic radiative potential (containing $\delta (r)$).

In neutral atoms there is an additional small parameter suppressing
the external photon vertex contribution (Fig. \ref{fig:rad_E1}(c)). The binding energies
of the valence electron and the external photon frequency are extremely small in
comparison with typical virtual photon frequencies. This makes corrections to the 
electric dipole operator (from the electron anomalous magnetic moment) and all contributions
proportional to $\frac{\partial \Sigma}{\partial E}$ negligible (see Eq. (\ref{semider})).

Therefore, the radiative potential approach should work in neutral
atoms even better than in hydrogen-like ions. One has only to add
the radiative potential to the atomic potential, calculate the electron
wave functions and use them to calculate electric dipole matrix elements.
It is convenient to perform the calculations in the length form for the electric dipole operator 
(${\bf D}=e{\bf r}$).

In the following section (Section \ref{low-energy_theorem}) we derive the low-energy theorem 
for the E1 amplitude (expressing the vertex contribution in terms of the self-energy). 
In Section \ref{E1_discussion} we explain the suppression of the vertex corrections 
and the validity of the radiative potential approach. 
In Section \ref{renorm_and_estimates} we perform a standard subtraction to remove ultraviolet 
divergences from the radiative corrections to the amplitude and estimate different contributions. 

\subsection{Derivation of the low-energy theorem}
\label{low-energy_theorem}

For the contribution of the high-frequency virtual photons
the low-energy theorem follows from the Ward identity \cite{berestetskii}
\begin{equation}\label{ward}
\Gamma^{\mu}(p,p,0)=\frac{\partial G^{-1}(p)}{\partial p_{\mu}} \ ,
\end{equation}
where $\Gamma^{\mu}$ is the vertex operator, $G^{-1}=\gamma^{\mu} p_{\mu}- m -M(p)$
is the inverse electron Green's function, and $M(p)$ is the mass
(self-energy) operator in the momentum representaion.
In the length form for the electric dipole operator we only
need the zero component of the Ward identity ($\Gamma^{0}=\gamma^{0}-\frac{\partial M(p)}{\partial E}$)
since the electron-photon interaction in this case is described by 
$H_{\rm int}\approx e \gamma ^{0} \Gamma^{0} \phi$, where $\phi ={\bf r\cdot  E}$ and ${\bf E}$ is 
the photon field. Therefore, the sum of the usual E1 amplitude and the vertex correction is 
\begin{equation}\label{rh}
<1|r - \frac12 (r \frac{\partial \Sigma}{\partial E} + \frac{\partial \Sigma}{\partial E} r)|2> \ ,
\end{equation}
where $\Sigma \equiv \gamma^{0}M$. 
Here it is assumed that the transformation of the mass operator from the momentum representation 
to the coordinate representation is accompanied by the antisymmetrization of the operators
$r$ and $M$ since, generally speaking, they do not commute (actually, this non-commutativity 
exists for the low-frequency contribution only). There is also some difference
in definitions of the mass operator $M$ in \cite{berestetskii} and the self-energy
operator $\Sigma$ used in this work (extra Dirac matrix $\gamma^{0}$;
the matrix elements in \cite{berestetskii} are defined using 
${\overline \psi}=\psi^{\dagger}\gamma^0$, we use $\psi ^{\dagger} \Sigma \psi$). 
Note that this low-energy theorem for the high-frequency contribution is valid in any
order of perturbation theory (including all orders in $Z \alpha$) and holds for the renormalized 
operators (similar to the Ward identity). One should add to Eq. (\ref{rh}) the QED corrections to 
the wave functions $|1>$ and $|2>$ which are not shown there explicitly.

To prove the low-energy theorem for the low-frequency contribution we can use
 a non-relativistic 
expression for the QED correction to the electric dipole amplitude presented
 e.g. in Ref. \cite{Pachucki}: 
\begin{equation}\label{pah}
\delta <1|z|2> = \frac{2 \alpha}{3 \pi m^2}\int_0^{\kappa}\omega
 d\omega {\rm Re}{f(\omega)} \ ,
\end{equation}
\begin{equation}\label{pahsig}
f(\omega)=\sum_{n,k}\frac{<1|z|n><n|p_i|k><k|p_i|2>}{(E_2-E_n)'(E_2-E_k-\omega)}
+\sum_{n,k}\frac{<1|p_i|n><n|p_i|k><k|z|2>}{(E_1-E_n-\omega)(E_1-E_k)'}
\end{equation}
\begin{equation}\label{pahv}
+\sum_{n,k}\frac{<1|p_i|n><n|z|k><k|p_i|2>}
{(E_1-E_n-\omega)(E_2-E_k-\omega)}
\end{equation}
\begin{equation}\label{pahn}
-\frac{<1|z|2>}{2}\sum_{n}(\frac{<1|p_i|n><n|p_i|1>}
{(E_1-E_n-\omega)^2}+\frac{<2|p_i|n><n|p_i|2>}
{(E_2-E_n-\omega)^2}) \ .
\end{equation}
The typical frequency $\omega$ of a virtual photon is large in comparison with the 
difference of excitation energies of the valence electron. 
%Therefore, the vertex contribution (Eq. (\ref{pahv})) and the normalization contribution
%(Eq. (\ref{pahn})) are small in comparison with the contribution of the correction to the 
%wave function (Eq. (\ref{pahsig})); the ratio is $\sim (E_1-E_n)/(E_1-E_n-\omega)$.
Therefore, we can simplify the the vertex contribution (Eq. (\ref{pahv})) by replacing $E_n$ by $E_k$, 
since $(E_k-E_n) \ll \omega$, and summing over $n$ (closure). 
We repeat this procedure by replacing $E_k$ by $E_n$ and summing over $k$.
The result for the vertex contribution can be presented in a symmetric form
(to cancel the first order correction in $(E_k-E_n)/\omega$ and the commutator $[z,p_i]$)
\begin{equation}\label{struct}
\frac{1}{2}\sum_{n}(\frac{<1|z p_i|n><n|p_i|2>}
{(E_2-E_n-\omega)^2}+\frac{ <1|p_i|n><n|p_i z|2>}
{(E_1-E_n-\omega)^2}) \ .
\end{equation}
Here we neglected the small difference $E_2-E_1$.
Now we see that the vertex contribution is indeed proportional to 
$z \frac{\partial \Sigma}{\partial E}$ ($\Sigma \propto
\frac{p_i|n><n|p_i}
{(E_1-E_n-\omega)}$,
 $\frac{\partial \Sigma}{\partial E_1}\propto -
\frac{p_i|n><n|p_i}
{(E_1-E_n-\omega)^2}$). 
Combining all terms and using the definition of $\Sigma$,
 \begin{equation}\label{sig}
\Sigma(E_1) = \frac{2 \alpha}{3 \pi m^2}\int_0^{\kappa}\omega
 d\omega {\rm Re}{\sum_{n}\frac{p_i|n><n|p_i}
{(E_1-E_n-\omega)}} \ ,
\end{equation}
we obtain the low-energy theorem for the radiative correction to the electromagnetic amplitude:
 \begin{equation}\label{rA}
<1|D|2>_{\rm rad}= \sum_n \frac{<1|D|n><n|\Sigma(E_2)|2>}{E_2-E_n}
+ \sum_n \frac{<1|\Sigma(E_1)|n><n|D|2>}{E_1-E_n}
\end{equation}
\begin{equation}\label{rsA}
 -\frac{1}{2}<1|D\frac{\partial \Sigma}{\partial E}+\frac{\partial \Sigma}{\partial E}D|2>
\end{equation}
\begin{equation}\label{rnA}
 +\frac{1}{2}<1|D|2>(<1|\frac{\partial \Sigma}{\partial E}|1>+
<2|\frac{\partial \Sigma}{\partial E}|2>) \ .
\end{equation}
Note that we can extend this derivation to include the high frequency contribution. 
We just have to use the relativistic radiation operator $\gamma^{\mu} \exp{(i{\bf k r})}$ 
instead of the momentum operator $p_i$. 
(One should start from the relativistic expression for the amplitude \cite{BSunpub,LF1974,BS1978} 
instead of Eqs. (\ref{pah})-(\ref{pahn}).)   

In this derivation we used the long-range character of the electric dipole operator ${\bf r}$. 
Indeed, we assumed that the matrix element $<n|z|k>$ in the sum over $n$ and $k$ is dominated
by states with $|E_n-E_k| \ll \omega$. This is certainly correct for the states located at 
distances comparable to the radius of the valence electron where $|E_n| \sim |E_k| \sim |E_1| \sim |E_2|$. 
%(it is convenient to consider the ``atom in a box'' problem where there is no continuum). 
For short-range operators (e.g. weak or hyperfine interactions) the contribution of states 
$|E_n| \sim |E_k| \gg |E_1| \sim |E_2|$ may be important.

\subsection{Enhancement of the self-energy contribution}
\label{E1_discussion}

%Let us come back to the low-energy theorem, Eqs. (\ref{rA},\ref{rsA},\ref{rnA}), to explain the 
%suppression of the vertex contribution (structural radiation), which can be approximately expressed 
%in terms of $\frac{\partial \Sigma}{\partial E}$, compared to the contribution of the electron 
%self-energy operator $\Sigma(r,r',E)$.
The contribution of the electron self-energy $\Sigma$ to the E1 amplitude (Eq. (\ref{rA})) is 
enhanced by the small energy denominator $E_{1,2}-E_n$ corresponding to the excitation of an external electron. 
The vertex (\ref{rsA}) and normalization (\ref{rnA}) contributions are not enhanced since 
$\frac{\partial \Sigma}{\partial E} \sim \Sigma/\omega$, where $\omega \gg E_{1,2}-E_n$ is a 
typical virtual photon frequency.
One may conclude that the vertex and normalization contributions are suppressed relative to the 
self-energy contribution by a small factor $ (E_{1,2}-E_n)/\omega$. 
Moreover, the vertex and normalization contributions usually have opposite signs and partially 
cancel each other. This is seen if we introduce complete sums $\sum_n |n><n|$ between the operators 
$D$ and $\frac{\partial \Sigma}{\partial E}$ in the first and second terms in Eq. (\ref{rsA}); 
the large diagonal contributions in these sums ($|n>=|1>$ in the first term and $|n>=|2>$ in the second) 
cancel exactly the normalization contribution Eq. (\ref{rnA}).

There is another way to explain the suppression of the vertex contribution.
The product $r\frac{\partial \Sigma(r,r',E)}{\partial E}$ is small everywhere inside a neutral atom.
The matrix element of $r$ typically comes from the distance $r \sim a_B$. 
At this distance the nuclear Coulomb field is screened, $Z_{eff} \sim 1$;
therefore, in this region $\frac{\partial \Sigma(r,r',E)}{\partial E}$ cannot have $Z^2$ enhancement. 
On the other hand, at small distances where the nuclear charge is unscreened, $r$ is small. 
This can also be seen from the vertex diagram itself where the operator $r$ is locked inside
the virtual photon loop located at a small distance from the nucleus.
There is no such suppression for the radiative correction to the electron wave function. 
The radiative potential changes the energy of the electron. 
This in turn changes the large distance asymptotics of the electron wave function 
and the matrix element of $r$. 

In the approximate expression for the radiative potential we use in this work 
$\frac{\partial \Sigma(r,r',E)}{\partial E}=0$ and so there are no vertex or normalization contributions. 
The first two terms in the radiative correction Eq. (\ref{rA}) can be presented in terms
of corrections to the wave functions produced by $\Sigma$:
\begin{equation}\label{psir}
|1_{\Sigma}> = \sum_n\frac{<1|\Sigma(E_1)|n><n|}{E_1-E_n} \ , \qquad 
|2_{\Sigma}> = \sum_n\frac{<2|\Sigma(E_2)|n><n|}{E_2-E_n} \ ,
\end{equation}
\begin{equation}\label{rpsi}
<1|D|2>_{\Sigma}=<1_{\Sigma}|D|2>+<1|D|2_{\Sigma}> \ .
\end{equation}
We have checked that our approximate expression for the radiative potential gives correct 
diagonal matrix elements of $\Sigma$ (radiative shifts). 
Here we need the non-diagonal matrix elements $<1,2|\Sigma|n>$. 
However, the main contribution to the matrix element of $D=er \sim e a_B$ is given by the 
low-energy states with $E_n \sim \hbar^2/(m a_B^2) \sim  E_{1,2} \ll \omega$, where $\omega$ is the 
virtual photon energy.
Therefore, our approximate radiative potential should give such matrix elements $<1,2|\Sigma|n>$ correctly. 
Note, once again, that this statement is incorrect for the radiative corrections to the matrix elements of 
short-range operators like the weak and hyperfine interactions. In this case the states with 
$E_n \sim \hbar c/r_{\rm typical} >\omega$ give a significant contribution and use of the radiative 
potential designed to give matrix elements between low-energy electron states is not justified.

The derivation (and result) of the low-energy theorem we have presented above is similar to that of
the low-energy theorem for correlation corrections to the electric dipole amplitude in our work \cite{corr}. 
The vertex (structural radiation) and normalization correlation corrections are also
proportional to $\frac{\partial \Sigma}{\partial E}$ and suppressed by a factor 
$E_{\rm valence}/E_{\rm core}\sim 1/10$ where $E_{\rm valence}$ and $E_{\rm core}$ are 
ionization energies for the valence and core electrons.
%We checked by direct numerical calculations that the vertex and normalization contributions
%are indeed small in the case of one electron above closed subshells.
It is interesting to note that for the correlation corrections the vertex and normalization contributions
were found to be numerically small for both long-range (electric dipole) and short-range 
(weak, hyperfine) operators.

There is also a certain similarity between this theorem and the low-energy theorem (the Low theorem) 
for bremsstrahlung (see, e.g., \cite{berestetskii}). The main radiation comes from the external particle 
ends of the scattering diagram and is expressed in terms of the elastic scattering amplitude. 
The structural radiation (from inside the scattering vertex) is small and is expressed in terms 
of the derivative of the elastic amplitude. In our case we consider the radiation of a weakly bound electron
($E<0$) which is not so different from the radiation of an unbound particle ($E>0$) if the energy $E$ 
is small. 

\subsection{Estimates of different QED corrections}
\label{renorm_and_estimates}

All terms in Eqs. (\ref{pah},\ref{pahv},\ref{pahsig},\ref{pahn}) and (\ref{rA},\ref{rsA},\ref{rnA}) 
are ultraviolet divergent as $\kappa$ tends to infinity. Therefore, we have to perform a subtraction 
of a standard counter term in the expression for the self-energy operator,
\begin{equation}\label{sub}
\frac{p_i|n><n|p_i}
{(E-E_n-\omega)}-\frac{p_i|n><n|p_i}
{(-\omega)}=\frac{p_i|n><n|p_i (E-E_n)}
{\omega (E-E_n-\omega)} \ ,
\end{equation}
to cancel the linear divergence in Eqs. (\ref{pah},\ref{pahsig}) and regroup other terms to cancel
the logarithmic divergences. After the subtraction and commutation $p_i (E_2-H)=-
[p_i,H]+(E_2-H)p_i=i \nabla_i (-e\Phi)+(E_2-H)p_i$ the first term in Eq. (\ref{pahsig}) can be 
transformed as follows (we use the operator form of Eq. (\ref{pahsig}) for brevity):
\begin{equation}\label{sub1}
\frac{z}{2\omega}\frac{1}{(E_2-H)'}\Big(
i \nabla_i (-e\Phi)\frac{1} {(E_2-H-\omega)}p_i
-p_i \frac{1}{(E_2-H-\omega)}i \nabla_i (-e\Phi) \Big)
\end{equation}
\begin{equation}\label{sub2}
+\frac{1}{2\omega}(z p_i -z|2><2|p_i) \frac{1}{(E_2-H-\omega)}p_i \ .
\end{equation}
The second term in Eq. (\ref{pahsig}) gives a similar contribution.

All terms containing $i \nabla_i (-e\Phi)$ are combined to give a low-frequency contribution 
to the radiative potential.
In particular, the large-$\omega$ contribution in Eq. (\ref{sub1}) gives a well-known local 
term in $\Sigma$
\begin{equation}\label{sigdelta}
\Sigma({\bf r},{\bf r}',E)=\frac{\alpha }{3\pi m^2}\ln{\frac{\kappa}{E_{\rm min}}}\,\,
\nabla^2 (-e\Phi (r)) \,\,\delta ({\bf r}-{\bf r'})
\end{equation}
which cancels the low-energy cut-off parameter $\kappa$ in the high-frequency
contribution proportional to $\ln{\frac{m}{\kappa}}$ (from the formfactor $f$; see \cite{B}).
Here $\nabla^2 (-e\Phi (r))= 4 \pi e^2 Z \delta ({\bf r})$.
It is easy to estimate the contribution of $\Sigma$ to the QED radiative corrections 
to $<s|z|p>$ (see Eq. (\ref{rA})).
The non-relativistic density of an $s$-wave valence electron is $\psi^2(0)\sim (Z_i+1)^2 Z/a_B^3$, 
where $Z_i$ is the ion charge ($Z_i=0$ for a neutral atom and $Z_i=Z-1$ for a hydrogen-like
ion), the energy $E_1 \sim (Z_i+1)^2 \alpha^2 mc^2$.
Therefore, the relative value of the QED correction produced by $\Sigma$ is 
$\sim <s|\Sigma|s>/E_1 \sim \alpha^3 Z^2 \ln{(1/\alpha^2 Z^2)}$.  
  
The terms in Eq. (\ref{sub2}) are combined with the vertex and normalization contributions 
to produce equations which do not contain ultraviolet divergences.
The terms proportional to $z p_i$ (see Eq. (\ref{sub2})) and $p_i z$ combined with the 
vertex contribution (\ref{pahv}) give the following QED radiative correction to $<1|z|2>$:
\begin{equation}\label{subv}
<1|z|2>_{\rm vertex}= -\frac{\alpha}{3 \pi m^2}
\sum_{n,k} \langle 1|p_i|n \rangle \langle n|z|k \rangle \langle k|p_i|2 \rangle \frac{E_n-E_1+E_k-E_2}
{E_n-E_1-E_k+E_2}\ln{|\frac{E_n-E_1}
{E_k-E_2}|} \ .
\end{equation}
The energy-dependent factor in this equation is always of the order of unity 
(it varies from 2 to $\ln{|\frac{E_n}{E_k}|}$). 
The ratio $v/c$ for a valence electron is $p/mc \sim (Z_i+1) \alpha$.
Therefore, the relative correction from the vertex term is $\sim (Z_i+1)^2\alpha^3$.
A more sophisticated estimate based on closure gives the same result:
$ \alpha<1|pzp|2>/m^2=$
$ \alpha<1|p^2z +zp^2|2>/m^2=$
$2 \alpha<1|(E_1+e\Phi)z +z(E_2+e\Phi)|2>/m$
$\sim  \alpha^2 (Z_i+1)/m $
$\sim  \alpha^3 (Z_i+1)^2 <1|z|2> $.
For hydrogen-like ions this correction is $Z^2\alpha^3$, comparable to the radiative potential 
contribution ($\sim \alpha^3 Z^2 \ln{(1/\alpha^2 Z^2)}$).
However, for neutral atoms this correction is $\sim \alpha^3$, i.e. it is extremely small.

The terms proportional to $z|2><2|p_i$ and $p_i|1><1|z$ combined with the normalization contribution 
(\ref{pahn}) give the following QED radiative correction to $<1|z|2>$:
\begin{equation}\label{subn}
<1|z|2>_{\rm norm}= -\frac{\alpha}{3\pi m^2} \langle 1|z|2 \rangle 
[\langle 1 |p^2| 1 \rangle + \langle 2 |p^2| 2 \rangle ] \ .
\end{equation}
%The typical virtual photon frequency in this integral is very small, $\omega \sim (E_1-E_2)$. 
%In other words, important wavelengths here are larger than $a_B$. This is the reason why 
%this contribution does not contain $Z^2$ enhancement in neutral atoms since such enhancement
%comes from small distances $r<a_B/Z$ where the Coulomb potential is unscreened. 
For a valence electron $p/m \sim (Z_i+1)\alpha c$, therefore the relative QED correction can be 
estimated as $\sim (Z_i+1)^2\alpha^3$, of the same order as the vertex contribution.

Finally, the contribution of the external photon polarization operator (Fig. \ref{fig:rad_E1}(f)) 
also does not have the $Z^2$ enhancement, since it comes from $r\sim a_B$.

Thus, the only important contribution in neutral atoms is that of the radiative potential.

\section{Applications to neutral cesium}
\label{Cs}

In this section we apply the radiative potential method to calculate radiative corrections 
to energy levels and electromagnetic amplitudes in the neutral cesium atom. 
We limit our consideration to the $s$ and $p$ levels of the external electron which are 
important for the parity violation calculation.
All calculations are performed taking into account finite nuclear size.

\subsection{Energies}

To calculate the radiative corrections to the energy levels we add the radiative potential 
to the nuclear Coulomb potential and calculate the self-consistent direct and exchange potentials
obtained using Dirac-Hartree-Fock (DHF) equations for the electron core. 
This DHF potential includes the potential $\delta V$ (``core relaxation'') which arises from the 
change in the core electron wave functions due to the radiative potential. 
Then we calculate the energy levels of the external electron in this DHF potential produced by the core electrons. 
The next step is to include the correlation corrections. 
It is convenient to calculate these corrections using the correlation potential method \cite{corr,dzuba89} 
which takes into account all second-order correlation corrections and three dominating series of 
higher-order diagrams (screening of the electron-electron interaction, the hole-particle interaction, and 
iteration of the correlation self-energy) to all orders in the residual 
Coulomb interaction.
The  non-local and energy-dependent correlation potential $\hat{\Sigma}^{corr}({\bf r},{\bf r}^\prime,E)$ 
is defined by the equation for the correlation correction to the electron energy
$\delta E_n^{corr}=\langle n|\hat{\Sigma}^{corr}({\bf r},{\bf r}^\prime,E)|n \rangle $; 
it is defined in an analogous way to the radiative potential, Eq. (\ref{L}). 
We add $\hat{\Sigma}^{corr}$ to the Dirac-Hartree-Fock potential to include it to all orders.
The results of calculations for the radiative corrections are presented in Table \ref{tab:b}.

\begin{table}[h]
\caption{Radiative corrections to ionization energies in Dirac-Hartree-Fock field without relaxation ((DHF)$_0$), 
with relaxation  included ((DHF)$_0+\delta V $), and with correlation corrections included 
((DHF)$_0+\delta V +\hat{\Sigma}^{corr} $); units cm$^{-1}$.}
\label{tab:b}
\begin{ruledtabular}
\begin{tabular}{c|ccccccccccc}
level & $6s_{1/2}$ & $7s_{1/2}$ & $6p_{1/2}$ & $7p_{1/2}$ & $8p_{1/2}$& $9p_{1/2}$ \\ 
\hline
(DHF)$_0$ & 15.5 & 4.3 & 0.2 & 0.07 & 0.03 & 0.02 \\
\hline
(DHF)$_0+\delta V$ & 15.9 & 4.3 & -0.8 & -0.3 & -0.1 & -0.07  \\
\hline
(DHF)$_0+\delta V +\hat{\Sigma}^{corr} $ & 17.6 & 4.1 & -0.4 & -0.1 & -0.05 & -0.03 \\
\end{tabular}
\end{ruledtabular}
\end{table}

It is seen that the many-body corrections change the result for the $s$-levels by $\sim 10\%$ and 
they change the sign and magnitude for $p$-levels. 
Our results for Uehling relaxation (not explicitly presented) are in perfect agreement with those of 
Ref. \cite{uehling_relax}. 

The results are in agreement with our previous calculations \cite{dzuba2002} and lie within the range 
spanned by model potential calculations of the $6s$ Lamb shift performed in 
Ref. \cite{Labzowsky} (from 15 to 27 cm$^{-1}$) and Ref. \cite{Sapirstein_energies} (from 13 to 23 cm$^{-1}$). 
However, the accuracy of our present work is higher, $\sim 1\%$ for $s$-levels.  

\subsection{E1 amplitudes}

We use a similar method as that for the energies to calculate the radiative corrections to the 
electromagnetic amplitudes between the $s$ and $p$ levels.
First, we calculate the external electron wave functions including the radiative potential and the core
relaxation $\delta V$. Then we use these wave functions to calculate the radiative corrections to the 
electromagnetic amplitudes in the Dirac-Hartree-Fock approximation. 
At the second step we calculate the effect of the electron core polarization by the photon electric
field using the time-dependent Hartree-Fock method. 
These core polarization corrections are often called the RPAE (random phase approximation with exchange) 
corrections.
At the final step we use the correlation potential method \cite{corr,dzuba89} to calculate the correlation 
corrections to the radiative corrections. In fact, to the required accuracy it is enougth to add 
$\hat{\Sigma}^{corr}$ to the Dirac-Hartree-Fock equations and calculate the external electron wave functions.
Other correlation corrections are proportional to $\frac{\partial \hat{\Sigma}^{corr}}{\partial E}$ and contribute
about $1\%$ only (the low-energy theorem; we mentioned this at the end of Section \ref{E1_discussion}). 
The results of the calculations for the radiative corrections are presented in Table \ref{tab:c}. 
Following Ref. \cite{Sapirstein1}, we present the results in terms of the dimensionless 
relative radiative corrections $R$ to the electromagnetic amplitude defined by the relation
\begin{equation}\label{R}
\langle s|r|p \rangle=  \langle s|r|p \rangle _0 (1+\frac{\alpha}{\pi}R_{sp}) \ .
\end{equation}

\begin{table}[h]
\caption{Relative radiative corrections $R_{sp}$ to the electromagnetic amplitudes 
$\langle s_{1/2}|r|p_{1/2} \rangle=  \langle s_{1/2}|r|p_{1/2}\rangle _0 (1+\frac{\alpha}{\pi}R_{sp})$ 
in Dirac-Hartree-Fock field with relaxation included (DHF), with RPAE corrections included (DHF+RPAE), 
and with correlation corrections included (DHF+RPAE+$\hat{\Sigma}^{corr} $).}
\label{tab:c}
\begin{ruledtabular}
\begin{tabular}{c|ccccccccccc}
Transition & $6s-6p$ &  $7p$ &  $8p$ & $9p$ & $7s-6p$ & $7p$ & $8p$   & $9p$ \\ 
\hline
DHF& 0.266 & -2.90 & -4.62 & -5.68 & -0.451 & 0.270 & -2.07 & -3.21 \\
\hline
DHF+RPAE & 0.286 & -4.39 & -11.9 & -29.7 & -0.432 & 0.270 & -2.20 & -3.60 \\
\hline
DHF+RPAE+$\hat{\Sigma}^{corr} $ & 0.265 & -2.91 & -6.25 & -10.3 & -0.340 & 0.231 & -1.60 & -2.52  \\
\end{tabular}
\end{ruledtabular}
\end{table}

Note that the radiative corrections for the $6s_{1/2}-7p_{1/2},8p_{1/2},9p_{1/2}$ amplitudes are large 
since these amplitudes are small and sensitive to any corrections to the DHF potential.

Unfortunately, we calculated but did not present the radiative corrections to the electromagnetic 
amplitudes in our previous paper \cite{dzuba2002}; 
we presented only their total contribution to the parity violating amplitude.
Anyway, the accuracy in our present work is higher ($\sim 1\%$).

In a recent work \cite{Sapirstein1}, direct calculations of the radiative corrections 
to electromagnetic amplitudes in the neutral alkali atoms were performed.
In particular, the radiative correction to the $6s-6p$ amplitude in Cs was calculated using the local 
Kohn-Sham potential. In this approach, the non-local exchange interaction is replaced by a semi-empirical 
local term that depends on the electron density. Ref. \cite{Sapirstein1} also does not take into account 
many-body corrections: core relaxation, RPAE, and correlation corrections.
 Nevertheless, $6s-6p$ is a large ``resonance'' amplitude, and the many-body
effects should not be very significant here. Indeed, our Dirac-Hartree-Fock
 value  $R=0.266$ is very close to the result 
 of Ref. \cite{Sapirstein1}: $R=0.261$.  We select our Dirac-Hartree-Fock
 value for comparison since it is the lowest-order approximation we use (no RPAE or correlation corrections) 
and most similar to that of Ref. \cite{Sapirstein1}. 
However, the very small difference between the results is accidental.
The result of Ref. \cite{Sapirstein1} does not include
 the Uehling potential contribution ($\sim -10 \%$), and one may also
expect some difference due to the core relaxation effect and the different
treatment of the exchange interaction. 
 
\section{Radiative corrections to the PNC amplitude in cesium}
\label{PNC}

Now we can calculate the QED radiative corrections to the PNC $6s-7s$ amplitude in Cs.
It is convenient to use the results of the sum-over-states approach of Ref. \cite{blundell90}:
\begin{eqnarray}\label{sumcs}
E_{PNC}&=& 
\sum_n \frac{\langle 7s|D|np\rangle \langle np|\hat{H}_{W}|6s\rangle}
{E_{6s}-E_{np}}+\frac{\langle 7s|\hat{H}_{W}|np\rangle \langle np|D|6s\rangle}
{E_{7s}-E_{np}} \nonumber \\
&=&(1.908-1.352-0.070-0.020+...)+(-1.493+0.120+0.010+0.003+...) \nonumber \\
&=&-0.894+... \ .
\end{eqnarray} 
The units are $iea_{B}(-Q_{W}/N)\times 10^{-11}$, where $Q_{W}$ is the nuclear weak charge
and $N$ is the number of neutrons. 98\% of the sum is given by the terms $n=6,7,8,9$ explicitly 
presented above. The result includes the many-body corrections to all matrix elements. 

\subsection{Contributions of energies and E1 amplitudes}

Now it is very easy to calculate the contributions to the PNC amplitude from 
radiative corrections to the energy intervals and electromagnetic amplitudes. 
Using the last line of Table \ref{tab:b} (with all many-body corrections included) 
we find that the radiative corrections to the energy intervals change 
the PNC amplitude by -0.33\%. Using the last line of Table \ref{tab:c} 
(with all many-body corrections included) we find that the radiative corrections to the E1 amplitudes 
$\langle6s,7s|D|np_{1/2}\rangle $ change the PNC amplitude by +0.42\%. 

Note that in Cs these two corrections nearly cancel each other, the sum of the two contributions
is $\Delta$=(0.42-0.33)\%=0.09\%. For the first time this cancellation was noted in our 
work \cite{dzuba2002} where these corrections in the DHF approximation were estimated to be
0.33\% (E1) and -0.29\% (energies).
We found that the difference between our old and new results is mainly because in \cite{dzuba2002} 
we used only 3 dominating terms in the sum (\ref{sumcs}) while in the present work we use 8 terms.
The fact that in \cite{dzuba2002} we used a different (and less accurate) radiative potential is 
not so significant. The contribution of the radiative corrections to the omitted terms $n>9$ in the 
sum (\ref{sumcs}) in the present work is estimated (using an asymptotic formula) to be $\sim$ 0.01\%.

To estimate the error we also performed another calculation.
We neglected the low-frequency contribution $\Phi_l(r)$ and set the coefficient $A(Z,r)=1$ in  $\Phi_f(r)$. 
This  variation changes the $s$-wave shifts by a few per cent only. However, the $p$-wave 
shifts change several times since they are small and sensitive to the low-frequency contribution $\Phi_l(r)$. 
In this case the E1 contribution is 0.44\%, the energy contribution is -0.35\%. 
However, the sum  $\Delta$= 0.09\% does not change.
Thus, the value of $\Delta$ is very stable and practically does not depend on the choice of the (short-range) 
radiative potential if this potential gives correct energy shifts. 
We estimate the uncertanty in $\Delta=0.09\%$ as $\sim 0.01\%$.

\subsection{Weak matrix elements}
\label{weak}

The self-energy and vertex QED radiative corrections to the weak matrix elements have been calculated in 
Refs. \cite{kf_prl_2002,k_jpb_2002,MST_PRL_2002,kf_jpb_2003,MST_PRA_2003,SPVC_PRA_2003}) using Coulomb wave functions. 
However, all neutral atom wave functions near the nucleus are proportional to the Coulomb wave functions 
since the screening of the nuclear Coulomb potential and the electron energy may be neglected here.
Therefore, the relative magnitude of the QED radiative corrections for an external electron in a neutral atom 
is the same for all weak matrix elements $\langle np_{1/2}|\hat{H}_{W}|n's\rangle$ and coincides with the 
Coulomb case for $n,n' \gg 1$.
Moreover, even the many-body corrections do not influence this statement since these corrections
are proportional to the weak matrix elements $\langle np_{1/2}|\hat{H}_{W}|n's\rangle$
and are multiplied by the same factor (equal to the relative QED correction for the weak matrix element) 
which does not depend on $n$ and $n'$.
This means that we can use the Coulomb results of 
Refs. \cite{kf_prl_2002,k_jpb_2002,MST_PRL_2002,kf_jpb_2003,MST_PRA_2003,SPVC_PRA_2003}
to find the contribution of the QED corrections to the weak matrix elements. 
The results of calculations for the self-energy and vertex contribution are the following (in \%):
-0.73(20) \cite{kf_prl_2002} (all orders in $Z \alpha$, using approximate relation);
-0.6 \cite{k_jpb_2002} (lowest order);
-0.9(1) \cite{k_jpb_2002,kf_jpb_2003} (lowest order and estimate of higher orders);
-0.85 \cite{MST_PRL_2002,MST_PRA_2003} (lowest order and higher orders in the logarithmic approximation); 
-0.815 \cite{SPVC_PRA_2003} (all orders $2s-2p$). 
Note that even for $n=2$ the binding energy to mass ratio is $E_n/mc^2=0.02$ and the relative 
radiative correction should be very close to the large $n$ limit (i.e. one may expect a difference
with the large $n$ limit result $\sim E_n/mc^2=2 \%$).
Based on these results we assume the correction equal to $(-0.82 \pm 0.03)\%$ which is in agreement with 
all calculations.

The Uehling potential contribution is easy to find and the error is negligible. 
According to Refs. \cite{johnson01,milstein,dzuba2002,kf_jpb_2002,kf_jpb_2003} the Uehling contribution to
the weak matrix element is 0.42\% (note that the Uehling contribution to the weak matrix element is practically
the same as its contribution to $E_{PNC}$ since the sum of the Uehling contributions from the energies
and electromagnetic amplitudes is about -0.01\%). 
The Wichmann-Kroll contribution is very small, -0.005\% \cite{dzuba2002,Shabaev}.

Thus, the sum of all radiative corrections to the weak matrix elements contributes 
$(-0.82+0.42-0.005)\% \approx (-0.41 \pm 0.03)\%$ to $E_{PNC}$.

\subsection{PNC amplitude}

The sum of all QED radiative corrections to $E_{PNC}$ is 
-0.33\% (energies) + 0.42\% (E1) - 0.41\% (weak) = $-0.32 \pm 0.03$\%.
The error $\sim 0.01\%$ in $\Delta=$-0.33\% (energies) + 0.42\% (E1)  gives a small
contribution to the error in $E_{PNC}$ if added in quadruture. 
Note that all three contributions (energies, $E1$, weak) are equally important here, 
and the E1 contribution is the largest.

Recently, calculations of radiative corrections to the PNC amplitude in Cs 
were performed in Ref. \cite{Shabaev} in an effective atomic potential.
The result is $(-0.27 \pm 0.03)\%$.
Our result $(-0.32 \pm 0.03)\%$ is slightly different possibly due to the many-body corrections 
which have not been calculated in the work \cite{Shabaev}. 
Note that the net effect of the many-body corrections in Cs is not very large because of the 
accidental cancellations of different many-body corrections.
One can see these cancellations between the RPAE and correlation contributions in Table \ref{tab:c} 
for the radiative corrections to the E1 amplitudes.
A strong accidental cancellation happens also for the main PNC amplitude where there are 4 large 
correlation corrections, up to $\pm$20\% each, and the sum of all correlation corrections is 2\% only 
\cite{dzuba89,dzuba2002}; such a cancellation does not take place, for example, in the PNC amplitude 
for Tl, where the total contribution of the many-body corrections is large. 

The many-body calculations of the PNC amplitude produced by the electron-nucleus weak interaction 
are described in detail in our review \cite{FG} where one can also find numerous references
(see also the original papers for calculations of atomic structure 
\cite{dzuba89,blundell90,kozlov01,dzuba2002} and Breit corrections 
\cite{derevianko2000,sushkov01,dzuba01,kozlov01}).
The contribution of the weak electron-electron interaction is very small. 
Within the standard model it is equal to 0.04\% \cite{SF,milstein}. 
Therefore, the PNC amplitude is proportional to the nuclear weak charge $Q_W$. 
The result, including the radiative correction  $(-0.32 \pm 0.03)\%$ calculated in the
present work, is the following:
\begin{equation}\label{EPNC}
E_{PNC}= -0.898 (1 \pm 0.5\%)\times 10^{-11} iea_{B}(-Q_{W}/N) \ .
\end{equation}
From the measurements of the PNC amplitude \cite{wood97} we obtain
\begin{equation}\label{Qw}
Q_W=-72.66(29)_{\rm exp}(36)_{\rm theor} \ .
\end{equation}
The difference with the standard model value $Q_W^{\rm SM}=-73.19(13)$ \cite{Rosner} is
\begin{equation}\label{DeltaQ}
Q_W-Q_W^{\rm SM}=0.53(48) \ ,
\end{equation}
adding the errors in quadrature.

\section{Summary and conclusion}
\label{conclusion}

We suggest to calculate radiative corrections to energy levels and electromagnetic amplitudes 
using the simple radiative potential 
\begin{equation}\label{Phif}
\Phi_{\rm rad}(r)=\Phi_U(r)+\Phi_g(r)+ \Phi_f(r) +\Phi_l(r) + \frac23 \Phi^{\rm simple}_{WC}(r) \ ,
\end{equation}
where $\Phi_U(r)$ is the Uehling potential (\ref{U1}), $\Phi_g(r)$
is the magnetic formfactor contribution (\ref{g}), $\Phi_f(r)$ is the 
electric formfactor contribution (\ref{f}),
$\Phi_l(r)$ is the low-frequency contribution (\ref{l}).
The simplified Wichmann-Kroll potential $\Phi^{\rm simple}_{WC}(r)$ (\ref{WC}) gives a very 
small contribution which  may be noticeable ($\sim 1 \%$) only for $Z>80$. 

The results obtained using this radiative potential are in good agreement (few percent) with the 
radiative corrections to the $s$, $p_{1/2}$ and $p_{3/2}$ energy levels
calculated in \cite{mohr} for the hydrogen-like ions.
The advantage of the radiative potential method is that it is very simple and can be used in 
many-electron atoms and molecules.

To calculate radiative corrections to electric dipole amplitudes we suggest use of the 
low-energy theorem derived in Section \ref{low-energy_theorem}. 
It is applicable because the ionization energy of an external (valence) electron $E_{\rm ex}$ 
and the external photon frequency $\omega_{\rm ex}$ are small in comparison with the typical 
frequency of a virtual photon $\omega_{\rm v}$; $E_{\rm ex},\omega_{\rm ex} \ll \omega_{\rm v}$. 
The vertex and normalization corrections are expressed in terms of the energy derivative of the 
electron self-energy operator. The dominating contribution is given by the corrections to the 
electron wave functions produced by the radiative potential. 
The relative contributions of the remaining corrections are small, $\sim 1/Z^{2}$ in neutral atoms 
and $\sim (Z_i+1)^2/Z^2$ in ions. 

The radiative potential method allows us to take into account many-body effects. 
First, we add $\Phi_{\rm rad}$ to the  Dirac-Hartree-Fock Hamiltonian (i.e., use the potential 
$V=V_{DHF}+ \Phi_{\rm rad}$) and calculate the new self-consistent field which includes a correction 
to the Dirac-Hartree-Fock potential $\delta V_{DHF}$ due to the change of the internal electron orbitals 
produced by $\Phi_{\rm rad}$ (the relaxation effect).
The relaxation effect is always significant. For a short-range potential $\Phi_{\rm rad}$ it is larger 
than the direct radiative shift for electron angular momenta $j>1/2$. Then we can use the new electron 
energy levels and DHF orbitals to calculate the correlation corrections applying the many-body theory 
methods which are described, for example, in Ref. \cite{dzuba2002}. 

We applied the radiative potential method to calculate the radiative corrections to energy levels and 
electromagnetic amplitudes in the neutral Cs atom and demonstrated the importance of many-body effects
in such calculations. The many-body effects change the $s$-level radiative shifts by $\sim$ 10\%, 
and they change the sign and magnitude of the $p$-level shifts.
Many-body effects in the radiative corrections to the electromagnetic amplitudes are also very significant. 
The RPAE (core polarization) corrections usually enhance the radiative correction. 
The effect is especially significant for the small amplitudes where we observed the RPAE enhancement 
up to 6 times. The correlation corrections usually act in the opposite direction and are equally significant. 

Finally, we calculated the contributions of the radiative energy shifts and radiative corrections for the 
electromagnetic amplitudes to the PNC $6s-7s$ amplitude in cesium. 
The radiative corrections to the weak matrix elements have been calculated in previous works. 
The sum of all QED radiative corrections to $E_{PNC}$ is
-0.35\% (energies) + 0.42\% (E1) - 0.41\% (weak) = $(-0.32 \pm 0.03)\%$.
Note that all three contributions are equally important here, and the E1 contribution is the largest. 
Using this radiative correction and previous many-body calculations we obtain the PNC amplitude
$E_{PNC}= -0.898 (1 \pm 0.5\%)\times 10^{-11} iea_{B}(-Q_{W}/N)$.
From the measurements of the PNC amplitude \cite{wood97} we extract the Cs weak charge
$Q_W=-72.66(29)_{\rm exp}(36)_{\rm theor}$.
The difference with the standard model value $Q_W^{\rm SM}=-73.19(13)$ \cite{Rosner} is
$Q_W-Q_W^{\rm SM}=0.53(48)$.

\acknowledgments

We are grateful to M. Kuchiev and A. Yelkhovsky for useful discussions. 
We are also grateful to M. Kuchiev for providing Mathematica code
for calculations of the relativistic Coulomb wave functions
and to V. Dzuba for an improved version of the Dzuba-Flambaum-Sushkov code
for atomic many-body calculations.
This work was supported by the Australian Research Council.
JG acknowledges support from an Avadh Bhatia Women's Fellowship and from 
Science and Engineering Research Canada while at University of Alberta. 
JG is grateful to University of Alberta for kind hospitality on a subsequent 
visit funded by a UNSW Faculty of Science grant.

%************************************************************


\begin{thebibliography}{20}
\bibitem{B} 
A (non-relativistic) relation between the infrared cut-off parameters introduced in 
\cite{berestetskii} is $\ln{\lambda}=\ln{2\kappa}-5/6$.
\end{thebibliography}
\end{document}